\documentstyle[12pt]{article}

\title{
Overlapping of the characteristic regions in the decay on
heterogeneous centers with equal number density
}

\author{V.Kurasov}

\date{Victor.Kurasov@pobox.spbu.ru}

\begin{document}

\maketitle

The problem of the metastable phase decay has
some specific solutions depended
on the content of the condensating system. Namely, when the system contains
only one type of heterogeneous centers one can successfully apply the
iteration
procedure \cite{Kuni}. When there are at least two types of heterogeneous
centers all attempts to apply  the iteration procedure \cite{Novoj}
in the general situation fail. Then one has to consider the
set of characteristic situations \cite{Multidec}. This case will
be the subject of our analysis.

In some situations of the condensation process one
can use the iteration procedure, but the reconsidered variant. In other
situations one has to apply another theoretical approaches \cite{Multidec}.
In  \cite{Multidec}   it   was   announced   that   the   mentioned
characteristic situations
cover all variants of the possible external conditions (and the content
of   condensated system). The concrete proof is very long and dull
and it is omitted in \cite{Multidec} due to lack of  volume.

Ordinary the total number of heterogeneous centers is one and the same
for all types of centers.
For example, the total number of the positive ions is ordinary equal to
the total number of the negative centers.
This condition simplify the analysis and leads only to two characteristic
situations.

These situations are quite analogous to those considered in the general
situation \cite{Multidec}. Moreover, the way of proof is also quite analogous.
That's why we restrict ourselves by the demonstration of the proof in this
situation.

To formulate two characteristic situations  we have to recall
some definitions used in \cite{Multidec}. We shall start with
fundamental characteristics
of the condensation process. We shall mark by lower indexes  $+$ and $-$
two types of heterogeneous centers. In the real situation of ions these
indexes can mark the positive and negative centers correspondingly.
The absence of index means that the value can be refered to both types
of heterogeneous centers.

To describe  the first characteristic situation we denote by $\Delta F$
the height of the activation barrier. Certainly it is the function of the
supersaturation $\zeta$ defined as
$$
\zeta = \frac{n}{n_{\infty}} -1
$$
where
$n$ is the molecules number density of the vapor and
$n_{\infty}$ is the molecules number density of the saturated vapor.
As far as we have two types of centers
we have two values of the activation barrier heights $\Delta_+ F$ and
$\Delta_- F$.

The first characteristic situation is the situation of the "strong
unsymmetry".
It is characterized by the small value of parameter inverse to
$$
\delta \Delta F \equiv  | \Delta_+ F - \Delta_- F |
$$

More rigorously speaking we have to mention that really the small value
of parameter
$$
\exp(-|\Delta_+ F - \Delta_- F | )
$$
is required.

The second characteristic situation is the situation of the "moderate
unsymmetry". To formulate it we have to introduce the values of
$$
\Gamma  \sim  - \zeta \frac{d \Delta F}{d \zeta}
$$
As far as there are two types of centers we have
$$
\Gamma_+  \sim - \zeta \frac{d \Delta_+ F}{d \zeta}
$$
and
$$
\Gamma_-  \sim  - \zeta \frac{d \Delta_- F}{d \zeta}
$$

The second characteristic is characterized by the small value of parameter
$$
\delta_r \Gamma
 =
\frac{|\Gamma_+ - \Gamma_-|}{\Gamma_+ + \Gamma_-}
$$

Here we shall show that these two situations exhaust all
possibilities.

The structure of consideration is the following:
\begin{itemize}

\item
In the first part we shall present general reasons which allow to hope
that the mentioned situation can cover all possibilities of experimental
conditions.

\item
In the second part we shall derive the required overlapping for several models
of heterogeneous centers. This illustrates the proof and gives the answer
in some limit situations which will be used later.

\item
In the third part  we shall give the derivation of the statement for the
case of ions. The way of derivation will be also spread to the general
case.

\end{itemize}

All definitions from  \cite{Multidec} are acceptable.
We shall follow the system of units and definitions  used in these papers.

\section{General remarks}

The difference between two types of heterogeneous centers is induced by
some abstract charge $q$. In the situation of ions it is a real electric
charge, in other situations it is the abstract charge.

Let us suppose that we can vary $q$ starting from the zero value\footnote{In
the case of ions there is an elementary charge of an electron but we use
this way only for  a formal derivation.}. When $q=0$ we have $\Delta_+
F = \Delta_- F$ and $\Gamma_+ = \Gamma_-$. When $q$ is increasing
then $\delta_r \Gamma$ and $\delta  \Delta F$ are increasing also.
Let us stop when
\begin{equation}\label{l}
\delta \Delta F = 1
\end{equation}
 We mark this value of $q$ as $q_h$.

Having introduced parameter
$$
\delta_r \Delta F \equiv
\frac{\delta \Delta F}{\Delta_+ F + \Delta_- F }
$$
one can note that
\begin{equation}\label{ll}
\delta_r \Delta F |_{q_h} \ll 1
\end{equation}

The activation barrier height $\Delta F$ can be presented as
$$
\Delta F = \int_{\zeta}^{\zeta_0} \frac{\Gamma}{\zeta} d \zeta
$$
where
$\zeta_0$ is the supersaturation when the activation barrier disappears.
One can present the last integral as
$$
\Delta F (\zeta) = \frac{\Gamma (\zeta ') }{\zeta ' } (\zeta_0 - \zeta)
$$
where
$\zeta '$ is some value between $\zeta$ and $\zeta_0$.
Then
$$
\Delta F = \frac{\Gamma (\zeta)}{\zeta} \Delta \zeta
$$
where
$$
\Delta \zeta  \equiv
(\zeta_0 - \zeta)
\frac{\Gamma (\zeta ')}{ \Gamma (\zeta)  }
\frac{ \zeta }{ \zeta ' }
$$

The value $\Delta  \zeta$  has  the  sense  of  the  characteristic
distance
from $\zeta$ until the value where
the essential activation barrier  disappears in comparison
with initial value (this
value isn't        $\zeta_0$ ).
The function $(\zeta_0 - \zeta) / \Delta \zeta$ is a smooth function of
$\zeta$.

Two types of heterogeneous centers induces two values $\Delta_+ \zeta$
and $\Delta_- \zeta$.

One can easily show that
\begin{equation} \label{tt}
\Delta_+ \zeta  |_{q<q_h}
\approx
\Delta_- \zeta  |_{q<q_h}
\end{equation}
We shall show the last estimate very qualitatively.

Really, the barrier character of condensation implies that
$$
\Delta_+ F \gg 1
$$
$$
\Delta_- F \gg 1
$$
This leads also to\footnote{It is necessary for continious description
of the nearcritical region.}
$$
\frac{d \Delta_+ F}{d \zeta} = \frac{\Gamma_+}{\zeta} \gg 1
$$
$$
\frac{d \Delta_- F}{d \zeta} = \frac{\Gamma_-}{\zeta} \gg 1
$$

Then the violence of the  required condition leads to the violence of (\ref{l}).
So, one can now directly see (\ref{tt}).

One can  keep in mind that the most sharp function of the supersaturation
is $\Gamma$ and the smooth functions are $\Delta F$ and $\Delta \zeta$.
Then it is reasonable to transform this picture into the dependence on
$q$. Namely, one can consider that for the given supersaturation $\zeta$
 the value $\Gamma$ is the sharp function of $q$ and the values $\Delta
F$ and $\Delta \zeta$ are more smooth functions.
This can not be rigorously proven
but seems to be a reliable qualitative
picture. In the second section the similar facts will be justified for
concrete types of heterogeneous centers.

The mentioned approximate coincidence of $\Delta_+ \zeta$ and $\Delta_- \zeta$
allows to introduce
$$
\Delta \zeta  \equiv \frac{1}{2} (\Delta_+ \zeta + \Delta_- \zeta)
$$
and approximately substitute $\Delta_+ \zeta$ and $\Delta_- \zeta$ by
$\Delta \zeta$.

Then one can  rather approximately  show that
\begin{equation} \label{vv}
| \Delta_+ F - \Delta_- F | \approx
| \frac{\Gamma_+}{\zeta_+} -   \frac{\Gamma_-}{\zeta_-} |
\Delta \zeta
\end{equation}
and as far as
$$
\Delta \zeta / \zeta \ll 1
$$
one can come to
$$
| \frac{\Gamma_+}{\zeta_+} -   \frac{\Gamma_-}{\zeta_-} |
\Delta \zeta
\approx
\frac{| \Gamma_+ - \Gamma_- | }{\zeta_p} \Delta \zeta
$$
where
$\zeta_p = (\zeta_+ + \zeta_-) / 2 $.
In the last relation one can take as $\zeta_p$ approximately both $\zeta_+$
 and $\zeta_-$ with a rather small relative error.

Now one can express $\Delta \zeta$ through $\Delta F$ and substitute
it into the last relation.  Then it comes to
$$
|\Delta F_+ - \Delta F_-| \approx \frac{|\Gamma_+ - \Gamma_-|}{\zeta_p}
\frac{\zeta_p \Delta_{\pm} F }{\Gamma_{\pm}}
 = \frac{|\Gamma_+ - \Gamma_-|}{\Gamma_{\pm}} \Delta_{\pm} F
$$

The last estimate solves the problem of overlapping of the mentioned regions.
Really, as far as $\Delta_{\pm} F \gg 1$ we see that at $q_h$ where $|\Delta_+
F - \Delta_- F| \sim 1$ the small value $|\Gamma_+ - \Gamma_-| / \Gamma_{\pm}$
is guaranteed.

We have to note that the validity of (\ref{vv}) is the matter of question.
Certainly, one can adopt
$\Delta_{\pm} F \sim \Gamma_{\pm} \frac{\Delta \zeta } {\zeta} $,
but when we coming to the difference $\Gamma_+ - \Gamma_-$ the relative error
increases many times. This disadvantage leads to some further remarks.

\section{Model systems}

Here we shall consider three simple models and show the necessary overlapping
directly. This will illustrate that the overlapping of the regions $|\Delta_+
F - \Delta_- F| \geq 1$ and $\delta_r \Gamma \ll 1$ is rather natural.

\subsection{ Pseudo homogeneous model}

Suppose that $R_{c\ +} = R_{c\ -} = R_{c \ q=0}$ where $R$ is the radius
of the embryo, index "c" denotes the critical embryo.  This corresponds
to the weak influence of the nuclei on the surface region of the nearcritical
embryo. Later we shall approximately suppose that $\nu_{c\ +} = \nu_{c\
-} = \nu_{c\ hom}$ where $\nu$ is the number of the molecules inside the
embryo and index $hom$ corresponds to the homogeneously formed embryo.

Later the  index $hom$ differs from the subscript $q=0$
 (Index "q=0" supposes
only that the  terms depended on the sign of the charge are put to
zero. All other terms depended on the absolute sign of the charge are
conserved.) The role of the terms depended on the sign of a charge is rather
small in comparison with the role of the terms depended on the absolute
value of a charge\footnote{The homogeneous nucleation rate isn't between
the rates of embryos formation on the "positive" and "negative"
heterogeneous centers. Both
positive and negative centers are the active centers of the condensation.}.
So, one can use the  decomposition starting from  $q=0$  when  all  terms
independent
on the charge sign taken directly into account.

Now we shall return to direct calculations.  For $\Gamma_+$ and $\Gamma_-$
we have
$$
\Gamma_+ \sim \nu_{hom} - \nu_{e \ +}
$$
$$
\Gamma_- \sim \nu_{hom} - \nu_{e \ -}
$$

Then
$$
\frac{\Gamma_+ - \Gamma_- }{\Gamma_{\pm}} =
\frac{\nu_{e\ -} - \nu_{e\ +} } { \nu_{hom} - \nu_{e\  \pm}}
$$

For $|\Delta_+ F - \Delta_- F|$ we have a very rough estimate
$$
| \Delta_+F - \Delta_- F  | \sim F_-(\nu_{e\ -}) - F_+ (\nu_{e\  +})
$$
All dependence on sign is in the last difference. We have to estimate
the last difference by the smoothest dependence. Let us take the homogeneous
dependence for this value. Then
$$
F_-(\nu_{e\ -}) - F_+ (\nu_{e\  +}) \sim
\frac{a}{3} (\nu_{e\ +}^{2/3} - \nu_{e\ -}^{2/3}) \equiv p
$$
where $a$ is the renormalized surface tension.

Having estimated
$\nu_{hom} - \nu_{e \pm} > k \nu_{e \pm}$
with some parameter $k$ close to $1$
one can use for $(\Gamma_+ - \Gamma_-) / \Gamma_{\pm}$  very rough
(and smooth) estimate
$$
\frac{
\Gamma_+ - \Gamma_-}{\Gamma_{\pm}} < \frac{\nu_{e \ -} - \nu_{e \ +}}
{ k \nu_{e\ \pm}} \equiv \delta
$$
If $\delta$ is small then
$$
p = \frac{a}{3} (\nu_{e\ +}^{2/3} (1 - (1 - k \delta)^{2/3} )
 \approx A \delta k
$$
where
$$
A \equiv \frac{a}{3} \frac{2}{3} \nu_{e \ +}^{2/3} \gg 1
$$
The property $A  k \gg 1$ guarantees that the regions $\delta \ll 1$ and $p\geq
1$ are overlapped.

\subsection{Linear approximation model}

We shall start from the rigorous formula
$$
\Gamma_{\pm} \sim \nu_{c \ \pm} - \nu_{e \ \pm}
$$
When $\zeta = \zeta_{0 \pm} $ the difference in the r.h.s. goes to zero.
Now we shall introduce approximation
$$
\Gamma_{\pm} \sim \gamma_{\pm} (\zeta - \zeta_{0 \pm})
$$
which implies that the  difference
$\nu_{c \ \pm} - \nu_{e \ \pm}$ is the linear function of the supersaturation.
This approximation  has to be valid at some effective supersaturations
which makes the main contribution into the activation barrier
height\footnote{Due to $\frac{d \Delta F}{d \zeta} \sim \zeta \Gamma$
one can imagine $\Delta F$ as the result of growth of $\Delta F(\zeta)$
where $\zeta$ is falling from $\zeta_0$ to $\zeta$.}.
As it will be seen in the next subsection this approximation isn't valid
when $\zeta$ is near $\zeta_0$.

Having integrated the suggested approximation we come to
$$
\Delta_{\pm} F = \frac{\gamma_{\pm}}{2} (\zeta - \zeta_{0\pm})^2
$$

For $q=q_h$ we have
$$
1 = |\frac{\gamma_+}{2}(\zeta-\zeta_{0\ +})^2-\frac{\gamma_-}{2}(\zeta-\zeta_{0\
-})^2  |
$$
or
$$
1 = \Delta_+ F | 1 - \frac{\gamma_-}{\gamma_+} (\frac{\zeta - \zeta_{0 \ -}}{\zeta
- \zeta_{0 \ +}})^2 |
$$
The value
$ (\zeta - \zeta_{0 \ -})/(\zeta
- \zeta_{0 \ +})$ has to be close to $1$
or
it has to be  $\Delta_+F - \Delta_- F \sim
\Delta_+ F \gg 1$ which solves the situation.

Then
$$
1 = \Delta_+ F | 1 - \frac{\gamma_-}{\gamma_+}|
$$
or
$$
1 \gg (\Delta_+ F)^{-1} = |1- \frac{\gamma_-}{\gamma_+}|
$$
It can be valid only when
$$
 |1- \frac{\gamma_-}{\gamma_+}| \ll 1
$$

One can see that
$$
|1- \frac{\gamma_-}{\gamma_+}|
\sim |\frac{\Gamma_+ - \Gamma_-}{\Gamma_{\pm}}|
$$
So,
$$
 |\frac{\Gamma_+ - \Gamma_-}{\Gamma_{\pm}}| \ll 1
$$
which shows the overlapping.

\subsection{Moderate behavior model}

Now we shall construct the general model corresponding to the formation
and disappearing of the metastable state.
 The simplest form of the free energy corresponding to the appearance
of the gap in the region of the small sizes is following
$$
F \sim - b (\nu - \nu_0) + c (\nu - \nu_0)^3
$$
where $\nu_0$ is the characteristic
value\footnote{Here we are  interested only in behavior
of $F$ near $\nu_0$ and the asymptotic behavior of $F$ isn't
essential (it is wrong).}.

Here $b$ plays the role of the chemical potential (or supersaturation)
and $c$ is some negative parameter
associated with the nuclei and  independent (weakly dependent)
on the supersaturation.

Denoting $\nu - \nu_0$ via $x$ one can get
$$
- x_e = x_c = (\frac{b}{3 |c|})^{1/2}
$$
$$
\Delta F = \frac{4}{3^{3/2}} \frac{b^{3/2}}{|c|^{1/2}}
$$
$$
\Gamma \sim \frac{d \Delta F}{d b} = 2 x_0 =
\nu_c
- \nu_e
$$
which confirms $\Gamma \sim \nu_c - \nu_e$ directly.

The value of $b$ is the variable, the value of $c$ is supposed to be
parameter.
Now it is clear that $\Gamma_+$ essentially differs from $\Gamma_-$ only
when $c_+$ essentially differs from $c_-$. But it means that $\Delta_+
F \sim |c|^{-1/2}$ essentially differs from $\Delta_- F$. As far as
$\Delta_{\pm}
F \gg 1$ we see that it means that $|\Delta_+F - \Delta_- F| \gg 1$. So
the overlapping here can be also observed.

\section{Real systems}

\subsection{Ions}

Now we shall investigate the case of ions. The free energy of the embryo
formation on the electric charge $q$ can be presented in  leading terms
as following
$$
F =  - b \nu  + a \nu^{2/3} + c \nu^{1/3} +
(c_2 + c_q ) \nu^{-1/3}
$$
where $b$ is
the excess of the chemical potential in a liquid phase,
$a$ is the renormalized surface tension, $c$ and $c_2$ are some coefficients
depended on the absolute value of the nuclei charge, $c_q$ is the coefficient
proportional to the charge and, thus, depended on the sign of the charge.

It is more convenient to use instead of $\nu$ the variable $\rho \equiv
\nu^{1/3}$ which leads to
$$
F = - b \rho^3 + a \rho^2 + c \rho + (c_2 + c_q) \rho^{-1}
$$
For the critical size one can get
$$
- 3 b \rho^4 + 2 a \rho^3  + c \rho^2 = c_2 + c_q
$$

We shall present the critical  characteristics  in the following form
$$
\nu_{c\ \pm} =  \nu_{c\ 0}  + \delta_{\pm} \nu_c
$$
$$
\nu_{e\ \pm} =  \nu_{e\ 0}  + \delta_{\pm} \nu_e
$$
$$
\rho_{c\ \pm} =  \rho_{c\ 0}  + \delta_{\pm} \rho_c
$$
$$
\rho_{e\ \pm} =  \rho_{e\ 0}  + \delta_{\pm} \rho_e
$$
where index $0$ marks the values when $c_q = 0$ (but $c$ and $c_2$ are
conserved).

For $\rho_0$ (both for critical and equilibrium values) we have
$$
- 3 b \rho_0^4 + 2 a \rho_0^3  + c \rho_0^2 = c_2
$$

Then for $\delta \rho$ (here will be $\delta_+ \rho = - \delta \rho
\equiv \delta \rho$ in the main order
for both critical and equilibrium values)
one can get
$$
- 3 b (\rho_0 + \delta \rho)^4 +
2 a (\rho_0 + \delta \rho)^3 +
c (\rho_0 + \delta \rho)^2 =
c_2 + c_q
$$
and in the main order
$$
\delta \rho =
\frac{c_q}{-12 b \rho_0^3 + 6 a \rho_0^2 + 2 c \rho_0}
$$

For the critical value of the free energy one can get
$$
F \approx F_0 (\rho_0) + F_0''(\rho_0) \frac{\delta \rho^2}{2} + c_q
(\rho_0+ \delta \rho)^{-1}
$$
where
$$
F_0 = - b \rho^3 + a \rho^2 + c \rho + c_2 \rho^{-1}
$$
$$
F_0'' = - 6 b \rho + 2 a + 2 c_2 \frac{1}{\rho_0^3}
$$

Then one can come to
$$
F_{c\ +} - F_{c \ -} = 2 c_q \rho_{0\ c}^{-1}
$$
$$
F_{e\ +} - F_{e \ -} = 2 c_q \rho_{0\ e}^{-1}
$$
and
$$ \Delta_+ F - \Delta_- F = 2 c_q (\frac{1}{\rho_{0\ c}}
+ \frac{1}{\rho_{0\ e}}   )
$$

Now we shall turn to get $\Gamma = \nu_c - \nu_e$.
In the main order $\Gamma_+ = \Gamma_- = \Gamma_0
= \nu_{c 0 } - \nu_{e 0}$.
In the first order
$$
\Gamma_{\pm} =  \Gamma_0  \pm  (3 \rho_c^2 \delta \rho_c
- 3 \rho_e^2 \delta \rho_e)
$$
and
$$
\frac{\Gamma_+ - \Gamma_-}{\Gamma_0} =
\frac{6 \rho_c^2 \delta \rho_c - 6 \rho_e^2 \delta \rho_e}
{\rho_c^3 - \rho_e^3}
$$
Then one can take for the last value the following
estimate\footnote{If we take here $\delta \rho_e$ instead $\delta \rho_c$ all
consideration cen be repeated even in details. Then we have to take $V=
V(\rho_{e 0}) \gg 1$.}
$$
\frac{\Gamma_+ - \Gamma_-}{\Gamma_0} \sim
\frac{6 \delta \rho_c}
{\rho_c}
$$

When $q \sim q_h$ we get
$$
1 = 2 c_q (\frac{1}{\rho_{0\ c}} - \frac{1}{\rho_{0\ e}})
\sim \frac{2 c_q}{\rho_{0\ e}}
$$
and we see that $c_q \gg 1$.

Then for
$
\frac{\Gamma_+ - \Gamma_-}{\Gamma_0} $
we can find
$$
\frac{\Gamma_+ - \Gamma_-}{\Gamma_0} \sim
\frac{3 \rho_{0\ e}}
{\rho_{0\ c} V(\rho_{c 0})}
$$
where function $V$ is given by
$$
V (\rho) = -12 b \rho^3 + 6 a \rho^2 + 2 c \rho
$$
As far as
$V (\rho_{c 0}) \gg 1$
we see that
$$
\frac{\Gamma_+ - \Gamma_-}{\Gamma_0} \ll 1
$$
which proves the overlapping.

\subsection{Generalization for the arbitrary system}

Now we can use the last constructions to investigate  more general
situation.

Suppose we have $F_+$, $F_-$  and $F_0$. Then
$$F_{\pm} = F_0 + \delta_{\pm} F $$
For the critical and equilibrium values we have
$$
\nu_{c,e\ \pm} = \nu_{c,e\ 0} + \delta_{\pm} \nu_{c,e}
$$

To find $\delta_{\pm} \nu_{c,e}$ one can use
$$
\frac{d F_{\pm}}{d \nu} = 0
$$
or
$$
\frac{d F_0}{d \nu} + \frac{d \delta_{\pm} F}{d\nu} = 0
$$
or
$$
\frac{d F_0}{d \nu}|_{\nu=\nu_0}
+ \frac{d}{d\nu} (\frac{d F_0}{d\nu})|_{\nu=\nu_0} (\nu - \nu_0)
+ \frac{d \delta_{\pm} F}{d\nu} = 0
$$
which gives
$$
\delta_{\pm} \nu_{c,e} =
\frac{
\frac{d \delta_{\pm} F}{d\nu}|_{\nu = \nu_{c,e\ 0}} }
{ \frac{d^2  F_0}{d\nu^2}|_{\nu = \nu_{c,e\ 0}}
}
 $$

Ordinary in the leading term  $\delta_{\pm} F = \pm \delta F$. Then
$$
\delta_{\pm} \nu_{c,e} = \pm
\frac{\frac{d \delta F}{d\nu}|_{\nu = \nu_{c,e,\ 0}} }
{ \frac{d^2  F_0}{d\nu^2}|_{\nu = \nu_{c,e,\ 0}}
}
 $$

Now we shall find
$ \Delta_+ F - \Delta_- F$.
For $F(\nu_c)$ we get
$$F(\nu_c) = F_0(\nu_{c\ 0}) +
\frac{1}{2} \frac{d^2 F_0}{d\nu^2} \delta \nu_c^2 +
\delta F|_{\nu = \nu_{0 c}}
$$
Then
$$
F_{c\ +} - F_{c\ -} =      \frac{1}{3}
\frac{d^3 F_0}{d \nu^3} \delta \nu_c^3 + 2 \delta F |_{\nu=\nu_{0 c}}
$$ and
$$
\Delta_+ F - \Delta_- F =
\frac{1}{3}
\frac{d^3 F_0}{d \nu^3} \delta \nu_c^3 + 2 \delta F |_{\nu=\nu_{0 c}} +
....|_{\nu=\nu_e}
$$
The first term  corresponds to the opportunity  missed in the previous section.
Here we sgall consider it more correctly.

When $q=q_h$ we get
$$
1 =
\frac{1}{3}
\frac{d^3 F_0}{d \nu^3} \delta \nu_c^3 + 2 \delta F |_{\nu=\nu_{0 c}} +
....|_{\nu=\nu_e}
$$
 or
\begin{equation} \label{ii}
\frac{1}{3}
\frac{d^3 F_0}{d \nu^3}
\frac{(\frac{d \delta F}{d\nu})^3}{(\frac{d^2 F_0}{d \nu^2})^3 }
+ 2 \delta F |_{\nu=\nu_0} +
....|_{\nu=\nu_e} = 1
\end{equation}

To find $\Gamma_{\pm}$ we shall use $\Gamma \approx \nu_c - \nu_e$.
Then
$$
\frac{\Gamma_+ - \Gamma_-}{\Gamma_0} =
2
\frac{\delta \nu_c - \delta \nu_e}{\nu_{c\ 0} - \nu_{e\ 0}}
$$
and
$$
\frac{\Gamma_+ - \Gamma_-}{\Gamma_0} \sim
\frac{\delta \nu_c}{\nu_{c\ 0} - \nu_{e\ 0}}
$$

For further analysis one can express $\frac{d^2 F}{d \nu^2}$ in terms of the
halfwidht $\Delta_c \nu$ of the nearcritical region.
Then
$$
\frac{d^2 F}{d \nu^2} \sim
\Delta_c \nu^{-2}
$$
where $\Delta_c \nu$ is the halfwidth of the nearcritical region.
Then                        (\ref{ii})
transforms into
$$
\frac{1}{3}
\frac{d^3 F_0}{d \nu^3}
(\frac{d \delta F}{d\nu}|_{\nu_{0\ c}})^3
(\Delta_c \nu )^6 + 2 \delta F |_{\nu=\nu_{0\ c}} +
...|_{\nu=\nu_{0\ e}} \sim 1
$$

One can easily            prove that in the l.h.s. of last relation there
is no compensation and get the estimates
\begin{equation} \label{iii}
\frac{1}{3}
\frac{d^3 F_0}{d \nu^3}
(\frac{d \delta F}{d\nu}|_{\nu_{0\ c}} )^3  (\Delta_c \nu )^6 \leq 1
\end{equation}
$$
  \delta F |_{\nu=\nu_{0\ c}} \leq 1
$$

Then as far as
$$
\frac{\Gamma_+ - \Gamma_- }{\Gamma_0} =
\frac{
\frac{d \delta F}{d\nu}|_{\nu_{0\ c}} (\Delta_c \nu )^2 }
{\nu_c - \nu_e}
$$
according\footnote{ If we choose another opportunity
$\delta F \sim 1$ then having adopted $\delta F \sim c_q / \nu^{-\alpha}$
with some parameter $\alpha$ we repeat the previous section.} to (\ref{iii})
one can see that
$$
\frac{\Gamma_+ - \Gamma_- }{\Gamma_0} \sim
\frac{
1 }
{\nu_c - \nu_e} (
\frac{3}{\frac{d^3 F_0}{d \nu^3}})^{1/3}
$$

Having used the estimate
$$
\frac{d^3 F_0}{d \nu^3} \sim
\frac{d^3 F_{hom}}{d \nu^3} \sim a \nu^{-7/3}
$$
one can get
$$
\frac{|\Gamma_+ -    \Gamma_- | }{\Gamma_0} \sim
\frac{\nu_c^{7/9}}{\nu_c - \nu_e}
$$
which solves the problem when the denominator isn't too small. But the
small value of denominator is already investigated in the section "Moderate
behavior model".

For $\delta F \sim 1$ we have
$$
|\frac{\Gamma_- - \Gamma_+}{\Gamma_0}
\sim
\frac{\Delta_c \nu}{\nu_c - \nu_e} \alpha \Delta F (\frac{1}{\nu_c}
- \frac{1}{\nu_e} )
$$
which solves as far as
here $\delta F \sim 1$ the problem when $\nu_c$ isn't very close to
$\nu_e$. But this situation is already investigated in the
section "Moderate behavior model".

The leading property which allows to justify all necessary estimates is
the fundamental condition $\nu_c \gg 1$ which is necessary for the thermodynamic
description of the embryo.

\end{document}